\documentclass[11pt,a4paper]{article}
\usepackage{jheppub}

\title{Domain wall brane in squared curvature gravity}

\author{Yu-Xiao Liu,}
\author{Yuan Zhong\footnote{The corresponding author.},}
\author{Zhen-Hua Zhao}
\author{and Hai-Tao Li}

\affiliation{Institute of Theoretical Physics, Lanzhou University,\\
Lanzhou 730000, People' s Republic of China}
\emailAdd{liuyx@lzu.edu.cn,zhongy2009@lzu.edu.cn,zhaozhh09@lzu.edu.cn,liht07@lzu.edu.cn}

\abstract{ We suggest a thick braneworld model in the squared curvature gravity
theory. Despite the appearance of higher order derivatives, the localization of
gravity and various bulk matter fields is shown to be possible. The existence of the
normalizable gravitational zero mode indicates that our four-dimensional gravity is
reproduced. In order to localize the chiral fermions on the brane, two types of
coupling between the fermions and the brane forming scalar is introduced. The first
coupling leads us to a Schr\"odinger equation with a volcano potential, and the
other a P\"oschl-Teller potential. In both cases, the zero mode exists only for the
left-hand fermions. Several massive KK states of the fermions can be trapped on the
brane, either as resonant states or as bound states.}

\keywords{Field Theories in Higher Dimensions, Large Extra
Dimensions}

\begin{document}
%\toccontinuoustrue
\maketitle  %IS IGNORED %%%%%%%%%%%
%\flushbottom %to make the text fill the height of the page.
%If two or more equations are short, they can also be written on a single line, separated
%     by \quad or \qquad.

%\paragraph %preferred over non-numbered sections.
\section{Introduction}
The suggestion that our world might be a four-dimensional domain
wall was proposed early in the 1980s~\cite{Rubakov1983,Visser1985},
but it wasn't until the localization of gravity on either
thin~\cite{Randall1999,Randall1999a} or
thick~\cite{Gremm2000a,Csaki2000a,DeWolfe2000a,Wang2002,Bazeia2009}
domain wall was shown to be possible, that the braneworld models
attracted renew attentions. More and more studies indicate that the
braneworld models might be effective in solving some long existed
problems, such as the hierarchy problem and the cosmological
constant
problem~\cite{Rubakov2001,Perez-Lorenzana2004,Maartens2010}. In this
paper, we focus mainly on the thick domain wall case. A
comprehensive review about thick brane models can be found
in~\cite{Dzhunushaliev2010a}.

Some questions needed be addressed in the thick braneworld models:
the stability of the solution and the possibility of localizing
matter fields as well as gravity on the brane. In the frame of
Einstein's general relativity, there are a lot of works which
devoted to answer these questions,
see~\cite{Gremm2000a,Csaki2000a,DeWolfe2000a,Bazeia2009} for the
localization of gravity,
and~\cite{Randjbar-Daemi2000,Randjbar-Daemi2000a,Kakushadze2000,Youm2000,Oda2001,Oda2001a,Ringeval2002,Gogberashvili2003,Koley2005,Gogberashvili2006a,Melfo2006,Silva-Marcos2007}
for trapping various kinds of bulk matter fields on both thin and
thick branes. For recent developments
see~\cite{Liu2008c,Liu2008a,Liu2008,Liu2009,Liu2009a,Guerrero2010,Liu2010,Liu2010a,GogberashviliHerrera-AguilarMalagon-Morejon2010}.

However, due to the fact that general relativity is not
renormalizable, we are suggested to consider the effects of higher
order curvature terms~\cite{Stelle1977}. Such terms also appear in
the low-energy effective action of string
theory~\cite{Vilkovisky1992}. To evade the appearance of spin-2
ghosts, and the Ostrogradski instability~\cite{Woodard2007}, the
higher order curvature terms usually are introduced as the
Gauss-Bonnet term or an arbitrary function of the curvature, namely,
$f(R)$. Both cases were applied to consider various of issues in
cosmology and higher energy physics, for details,
see~\cite{Sotiriou2010,DeFelice2010,Nojiri2010} and references
therein.

As to the braneworld scenarios, it is shown in many works that the
introduce of the Gauss-Bonnet term usually imposes no impact on the
localization of
gravity~\cite{Corradini2000,Giovannini2001,Nojiri2001,Neupane2001,Germani2002,Lidsey2003,Davis2003,Cho2003,Sami2004,Barrab`es2005}.
This result might change in $f(R)$ gravity, however, because now
some higher order derivatives evolve in. The appearance of higher
order derivatives usually renders the constructing of a braneworld
models very hard.

For pure gravitational systems which contains no any matter fields,
some thin braneworld models have been constructed in a lower order
frame~\cite{Parry2005,Bronnikov2007,Deruelle2008,Balcerzak2008} by
using the Barrow-Cotsakis theorem~\cite{Barrow1988}. This theorem
states that the fourth order $f(R)$ gravity (higher order frame) is
conformally equivalent to a second-order gravity theory (lower order
frame). Unfortunately, for thick domain wall models, this method
would lead to ambiguous~\cite{Barvinsky2008} and thus is not a good
choice.

Some thick brane solutions were found directly in the higher order
frame~\cite{Afonso2007,Dzhunushaliev2010}. In~\cite{Afonso2007},
with a background scalar field, the authors found some analytical
thick brane solutions in both constant and variant curvature cases.
While in~\cite{Dzhunushaliev2010}, by analyzing the existence of the
fix points, the authors found some numerical thick $f(R)$-brane
solutions with pure gravity. The trapping of complex scalar field on
the brane solutions was shown to be possible
in~\cite{Dzhunushaliev2010}. Both papers considered the case where
$f(R)\propto R^n$.

However, the solutions given in~\cite{Afonso2007,Dzhunushaliev2010}
are not perfect for the following reasons:
\begin{itemize}
  \item The solution in
$M_5$ space given in~\cite{Afonso2007} contains singularity, because
the warp factor is not smooth at the brane. Meanwhile, even though
the solutions found in both $dS_5$ and $AdS_5$ spaces are stable
under the tensor perturbations, only the solution in $dS_5$ space
supports normalizable zero gravitational mode, see~\cite{Zhong2011}.
Therefore, the gravity can be localized only on the brane in $dS_5$
space. However, the constant curvature space is a rather special
case.
\item For the case of non-constant curvature, some solutions were
suggested in~\cite{Afonso2007,Dzhunushaliev2010}. However, these
solutions are not very good: the solution in~\cite{Afonso2007}
contains a singular point at where the curvature diverges; while the
solution in~\cite{Dzhunushaliev2010} is given numerically, and we
only know its behavior in infinity.
  \item Besides, the model
in~\cite{Dzhunushaliev2010} is purely gravitational, namely, no
matter field were introduced. However, to localize fermions, one
usually needs to couple the fermion with a scalar field.
\end{itemize}

 For the reasons above, we aim at
constructing a domain wall brane which has the following properties:
\begin{itemize}
  \item The brane solution is found in a system which contains higher order curvature terms and especially a system contains fourth-order derivatives.
  \item The brane is smooth, stable and is formed by a background scalar field.
  \item The gravity can be localized on such brane.
  \item By introducing the Yukawa coupling between the fermion and the background scalar field, the fermions can be trapped on the brane.
\end{itemize}

 As a toy model, let us focus on the squared
curvature gravity, i.e., the higher curvature correction is given by
$f(R)\propto R^2$. For this special gravity, a thin braneworld has
been constructed in~\cite{Parry2005} by using the Barrow-Cotsakis
theorem.

In the next section, we give a thick domain wall solution in the
squared curvature gravity. In section~\ref{section3}, we discuss the
localization of four-dimensional gravity. In section~\ref{section4},
the trapping of bulk matter fields with various spins is analyzed,
but we mainly focus on the half spin fermions. Our conclusions are
listed in the last section.
\section{Model and solution}
\label{section2}

We start with the action
\begin{eqnarray}
  S=\int d^5x\sqrt {-g}\left(\frac{1}{2\kappa_5^2}f(R)
  -\frac12\partial^M\phi\partial_M\phi-V(\phi)\right),
  \label{action}
\end{eqnarray} where $\kappa_5^2=8\pi G_5$ is the gravitational coupling constant
with $G_5$ the five-dimensional Newtonian constant. We have
introduced a background scalar field $\phi$, of which the
self-interacting is described by the potential $V(\phi)$. We prefer
to use Roman letters to denote the bulk coordinates, i.e.,
$M,N=0,1,2,3,4$ and the Greek letters represent the brane
coordinates, i.e. $\mu,\nu=0,1,2,3$.

Our discussions will be limited on the static flat braneworld
scenario, for which the metric is given by
\begin{eqnarray}
  ds^2=e^{2A(y)}\eta_{\mu\nu}dx^\mu dx^\nu+dy^2,
  \label{metric}
\end{eqnarray}
with $y=x^4$ the extra dimension. Meanwhile, the scalar field,
$\phi=\phi(y)$, is independent of the brane coordinates. For
system (\ref{action})-(\ref{metric}), the Einstein equations read
\begin{eqnarray}
\label{EE1}
  f(R)+2f_R\left(4A'^2+A''\right)
  -6f'_RA'-2f''_R=\kappa_5^2(\phi'^2+2V),
\end{eqnarray}
and
\begin{eqnarray}
\label{EE2}
  -8f_R\left(A''+A'^2\right)+8f'_RA'
  -f(R)=\kappa_5^2(\phi'^2-2V),
\end{eqnarray} where the primes represent derivatives with respect to
the coordinate $y$ and $f_R \equiv df(R)/dR$. The scalar field is described
by the following equation:
\begin{eqnarray}
 4A'\phi'+\phi''-\frac{\partial V}{\partial\phi}=0.
 \label{Eom}
\end{eqnarray}

We consider a toy model, in which $f(R)=R+\gamma R^2$. Then we would
have $f_R=1+2\gamma R,~f_R'=2\gamma R',~f_R''=2\gamma R'',~R'=
-(8A'''+40A'A''),~R''=-(8A''''+40A''^2+40A'A''')$. When $\gamma=0$,
we come back to the case of general relativity. Obviously, if
$\gamma\neq0$, the Einstein equations are fourth-order differential
equations. In order to find a domain wall solution, let us consider
the following $\phi^4$ model:
\begin{eqnarray}
V(\phi)=\lambda^{(5)}(\phi^2-v^2)^2+\Lambda_5,
\label{scalarPotential}
\end{eqnarray} where $\lambda^{(5)}>0$ is the self coupling constant of the scalar field,
and $\Lambda_5$ is another constant. Obviously, at $\phi=\pm v$ the
scalar potential $V(\phi)$ takes its minimum value, i.e.,
$V(\phi=\pm v)=\Lambda_5$.

It is easy to proof that eq.~\eqref{Eom} supports the following
solution:
\begin{eqnarray}
\label{solutionpart1}
\phi (y)&=&v \tanh \left(\sqrt{\frac{2\lambda ^{(5)}}{3}} v y\right),\\
\label{solutionpart2}
e^{A(y)}&=&\cosh^{-1}\left(\sqrt{\frac{2\lambda ^{(5)}}{3}} v y
\right).
\end{eqnarray}
However, the Einstein equations \eqref{EE1}-\eqref{EE2} strongly
constrain the possible values of the parameters. Specifically, the
Einstein equations are satisfied only when the parameters take the
following values:
\begin{eqnarray}
\label{solutionpart3} \lambda ^{(5)}=\frac3{784} \frac{\kappa
_5^2}{\gamma} ,\quad v= 7 \sqrt{\frac{3}{29\kappa _5^2}},\quad
\Lambda_5=-\frac{477}{6728 }\frac{1} {\gamma\kappa _5^2}.
\end{eqnarray}
Defining a new parameter by
\begin{eqnarray}
k=\sqrt{\frac{3}{232 \gamma }},
\end{eqnarray} we can rewrite
(\ref{solutionpart1})-(\ref{solutionpart2}) in a more simpler form:
\begin{eqnarray}
\phi (y)=v \tanh(k y),\quad e^{A(y)}=\cosh(ky)^{-1}.
\end{eqnarray}
For this reason, in the discussions below,
we always use the parameter $k$ rather than $\gamma$, to reflect the influence from the squared curvature term.% If we

\begin{figure}
%\begin{center}
\includegraphics[width=0.5\textwidth]{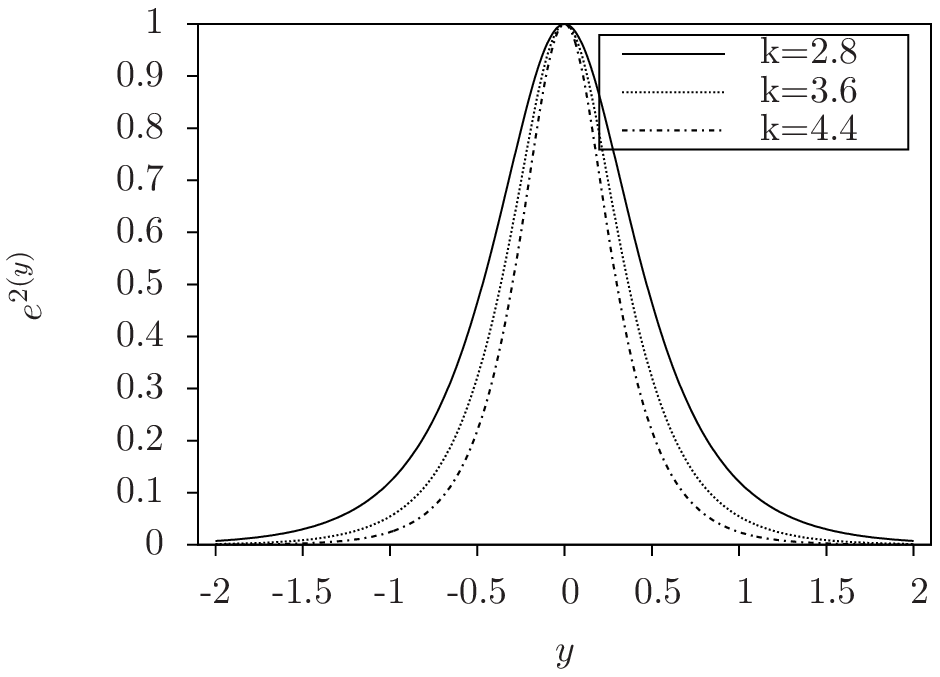}
\includegraphics[width=0.5\textwidth]{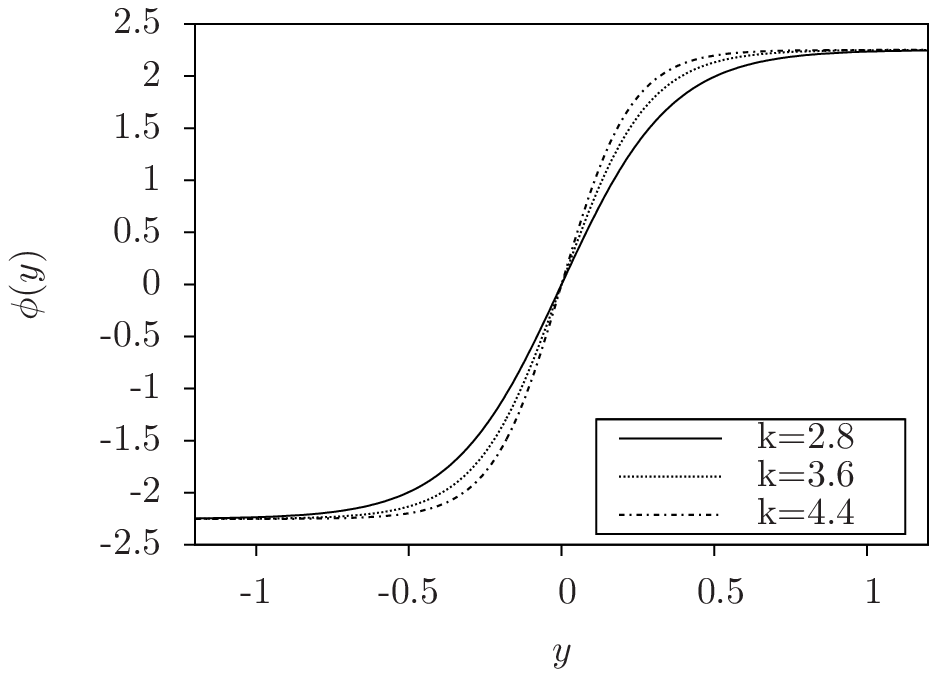}
\includegraphics[width=0.5\textwidth]{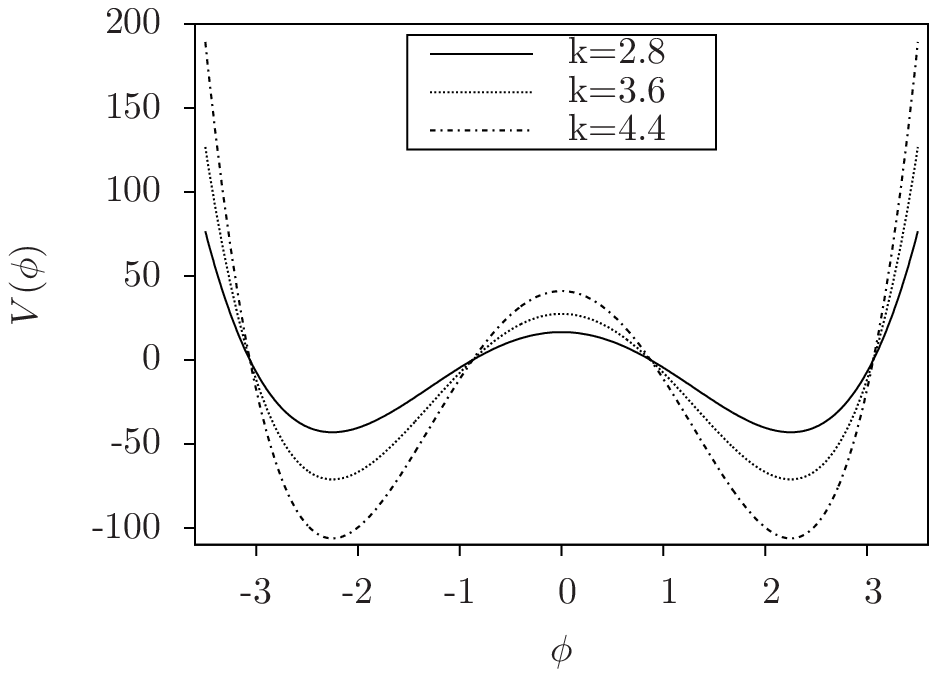}
\includegraphics[width=0.5\textwidth]{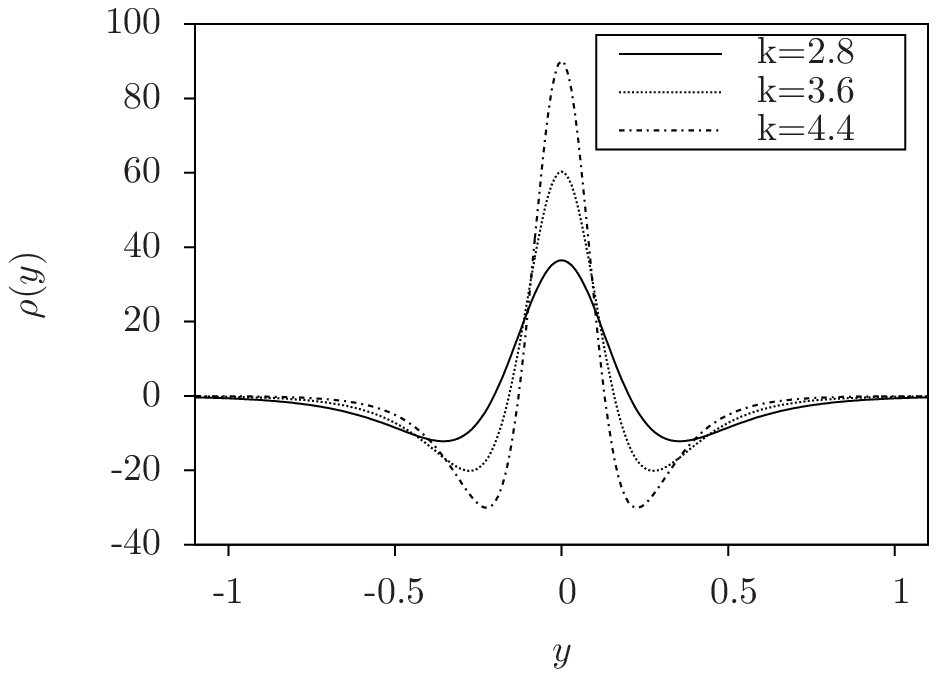}
%\end{center}
\caption{The dependence of the domain wall properties on the
parameter $k$.} \label{figureProperties}
\end{figure}

As shown in figure~\ref{figureProperties}, the scalar field is a
kink which satisfies $\phi(0)=0,~\phi(\pm\infty)=\pm v.$ $V(\phi)$
takes its minimum at $\phi=\pm v.$ We also see how the energy
density $\rho=T_{00}=e^{2A}(\frac12\phi'^2+V(\phi))$ distributes
along the fifth dimension. Obviously, $\rho$ peaks at $y=0$, where
the brane locates at. It is not difficult to verify that the
geometry of the space-time becomes anti-de Sitter at $y=\pm\infty$,
where the bulk curvature $R=-20k^2$ and the corresponding
cosmological constant is nothing but the parameter $\Lambda_5$. The
value of $\Lambda_5$ can also be obtained from eq.~\eqref{EE1} by
taking the limit $y\to\infty$.
%\begin{eqnarray}
%T_{00}&=&e^{2A}(\frac12\phi'^2+V(\phi))\nonumber\\ &=&\Lambda _5~
%\textrm{sech}^2\left(\sqrt{\frac{2\lambda^{(5)}}{3}} v y
%\right)+\frac{4}{3} v^4 \lambda^{(5)}
%\textrm{sech}^6\left(\sqrt{\frac{2\lambda^{(5)}}{3}} v y \right),
%\end{eqnarray}

\section{Tensor perturbations and the localization of gravity}
\label{section3}

To discuss the localization of massless four-dimensional gravity, we
need to consider the tensor perturbations:
\begin{eqnarray}
ds^2=e^{2A(y)}(\eta_{\mu\nu}+h_{\mu\nu})dx^\mu dx^\nu+dy^2,
\end{eqnarray}
where $h_{\mu\nu}=h_{\mu\nu}(x^{\rho},~y)$ depend on all the
coordinates and satisfy the transverse-traceless condition
\begin{eqnarray}
 \eta^{\mu\nu}h_{\mu\nu}=0=\partial_\mu h^\mu_{~\nu}. \label{TT}
\end{eqnarray}
The perturbation of the scalar field is denoted as
$\delta\phi=\tilde{\phi}(x^{\mu},~y)$.

In~\cite{Zhong2011}, we have demonstrated that the scalar
perturbation $\tilde{\phi}$ decouples from the tensor perturbations
$h_{\mu\nu}$ under the transverse-traceless condition (\ref{TT}),
and the perturbed Einstein equations are given by
\begin{eqnarray}
\left(\partial_y^2+4A'\partial_y+e^{-2A}\square^{(4)}\right)h_{\mu\nu}=-\frac{f_R'}{f_R}\partial_y h_{\mu\nu}~,
\end{eqnarray}
or in a more simpler form
\begin{eqnarray}
\square^{(5)}h_{\mu\nu}=\frac{f_R'}{f_R}\partial_y h_{\mu\nu}~.
\label{perturbationeq}
\end{eqnarray}
Here $\square^{(4)}=\eta^{\mu\nu}\partial_{\mu}\partial_{\nu}$ and
$\square^{(5)}=g^{MN}\nabla_{M}\nabla_{N}$ are the four-dimensional
and five-dimensional d'Alembert operators, respectively.

Obviously, for $R=\rm{const.}$ or $f(R)=R$, which simply gives the
theory of general relativity, eq.~(\ref{perturbationeq})
reduces to the five-dimensional Klein-Gorden equation for the
massless spin-2 gravitons, and the reproduction of brane gravity has
been discussed in~\cite{Gremm2000a,Csaki2000a,DeWolfe2000a}.
However, for an arbitrary form of $f(R)$ with variant $R$, the issue
of localization of gravity should be given a further consideration.

Following the method given
in~\cite{Gremm2000a,Csaki2000a,DeWolfe2000a}, we introduce a
coordinate transformation $dz=e^{-A(y)}dy$, under which the metric
(\ref{metric}) is conformally flat:
\begin{eqnarray}
  ds^2=e^{2A(z)}(\eta_{\mu\nu}dx^\mu dx^\nu+dz^2).
  \label{metricz}
\end{eqnarray}
It is very easy to proof that in the new coordinates system,
$y=\textrm{arcsinh}(k z)/k$ and therefore the warp factor takes the
simple form:
\begin{eqnarray}
  e^{2A(z)}=(1+k^2 z^2)^{-1}.
  \label{A_z}
\end{eqnarray}
Then the perturbed equation
(\ref{perturbationeq}) reads
\begin{eqnarray}
 \left[\partial_z^{~2}
  +\left(3\frac{\partial_z a}{a}
  +\frac{\partial_z f_R}{f_R}\right)\partial_z
  +\square^{(4)}\right]h_{\mu\nu}=0.
\end{eqnarray} Here we have denoted $a(z)=e^{A(z)}$ to write the final result looks more
symmetrical.

By doing the decomposition
$h_{\mu\nu}(x^{\rho},z)=(a^{-3/2}f_R^{-1/2})\epsilon_{\mu\nu}(x^{\rho})\psi(z)$,
with $\epsilon_{\mu\nu}(x^{\rho})$ satisfying the
transverse-traceless condition
$\eta^{\mu\nu}\epsilon_{\mu\nu}=0=\partial_\mu \epsilon^\mu_{~\nu}$,
we obtain a Schrodinger like equation for $\psi(z)$:
\begin{eqnarray}
  \left[\partial_z^2
      -W(z)\right]\psi(z)
      =-m^2\psi(z),\label{Schrodinger}
\end{eqnarray} with the potential given by~\cite{Zhong2011}
\begin{eqnarray}
 W(z)=\frac34\frac{a'^2}{a^2}
      +\frac32\frac{a''}{a}
      +\frac32\frac{a' f_R'}{a f_R}
      -\frac14\frac{f_R'^2}{f_R^2}
      +\frac12\frac{f_R''}{f_R}.
      \label{Schrodingerpotential}
\end{eqnarray}
One can also factorize the Schrodinger equation (\ref{Schrodinger})
as
\begin{eqnarray}
 \left[\left(\partial _z
 +\left(\frac{3}{2}\frac{\partial_z a}{a}+\frac{1}{2}\frac{\partial_z f_R}{f_R}\right)\right)
 \left(\partial_z
 -\left(\frac{3}{2}\frac{\partial_z a}{a}+\frac{1}{2}\frac{\partial_z f_R}{f_R}\right)\right)\right]\psi(z)
 =-m^2\psi(z)
\end{eqnarray} to conclude that there
is no gravitational mode with $m^2<0$ and therefore the solution in
(\ref{solutionpart1})-(\ref{solutionpart3}) is stable. For eq.
(\ref{Schrodinger}), the zero mode (for which $m^2=0$) reads
\begin{eqnarray}
\psi^{(0)}(z)=N_0 a^{3/2}(z)f_R^{1/2}(z), \label{zeromode}
\end{eqnarray}
where $N_0$ represents the normalization constant.

\begin{figure}
\begin{center}
\includegraphics[]{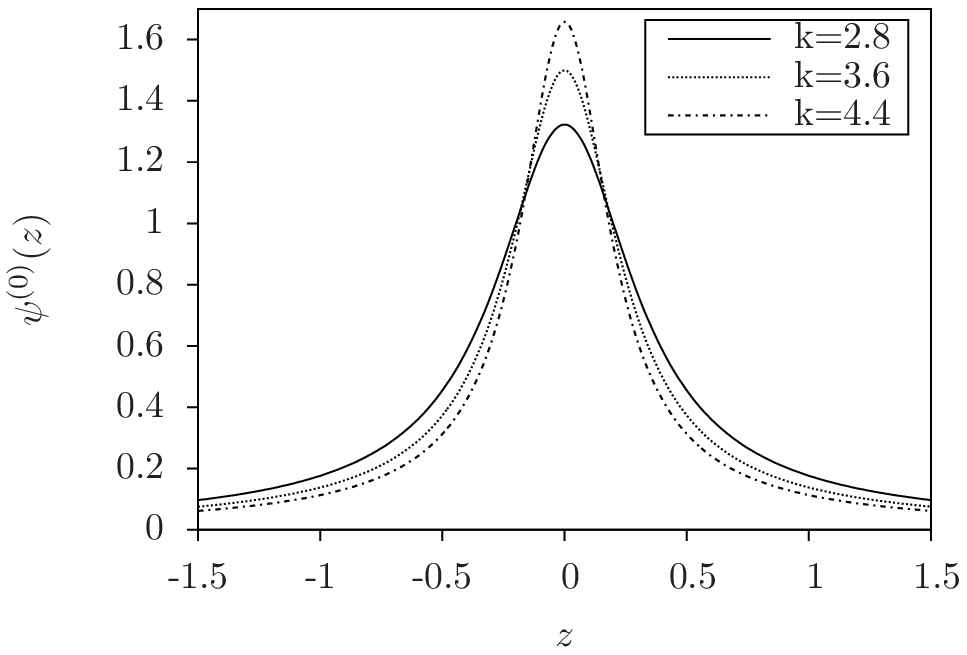}
\end{center}
\caption{The shapes of the gravitational zero mode $\psi^{(0)}(z)$.}
\label{figureGravZeroMode}
\end{figure}

By plugging (\ref{A_z}) into (\ref{Schrodingerpotential}), one
obtains
\begin{eqnarray}
W(z)=\frac{15 k^2 \left(-14+37 k^2 z^2+28 k^4 z^4+4 k^6
z^6\right)}{4 \left(5+7 k^2 z^2+2 k^4 z^4\right)^2},\label{3.11}
\end{eqnarray} the famous volcano
potential~\cite{Randall1999,Gremm2000a,DeWolfe2000a,Csaki2000a}.
This potential supports a normalizable zero mode (see
figure~\ref{figureGravZeroMode})
\begin{eqnarray}
    \psi^{(0)}(z)
    =\sqrt{\frac{k}{8}}\frac{\sqrt{5+2 k^2z^2}}{ \left(1+k^2
    z^2\right)^{5/4}},
\end{eqnarray} as well as a series of continuous massive KK
modes.

\begin{figure}
%\begin{center}
\includegraphics[width=0.5\textwidth]{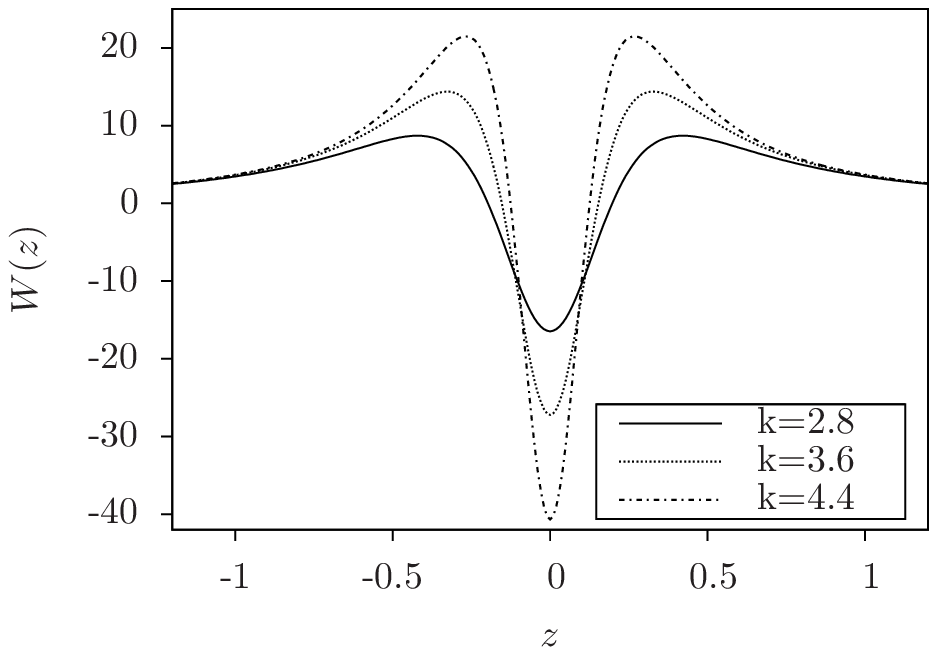}
\includegraphics[width=0.5\textwidth]{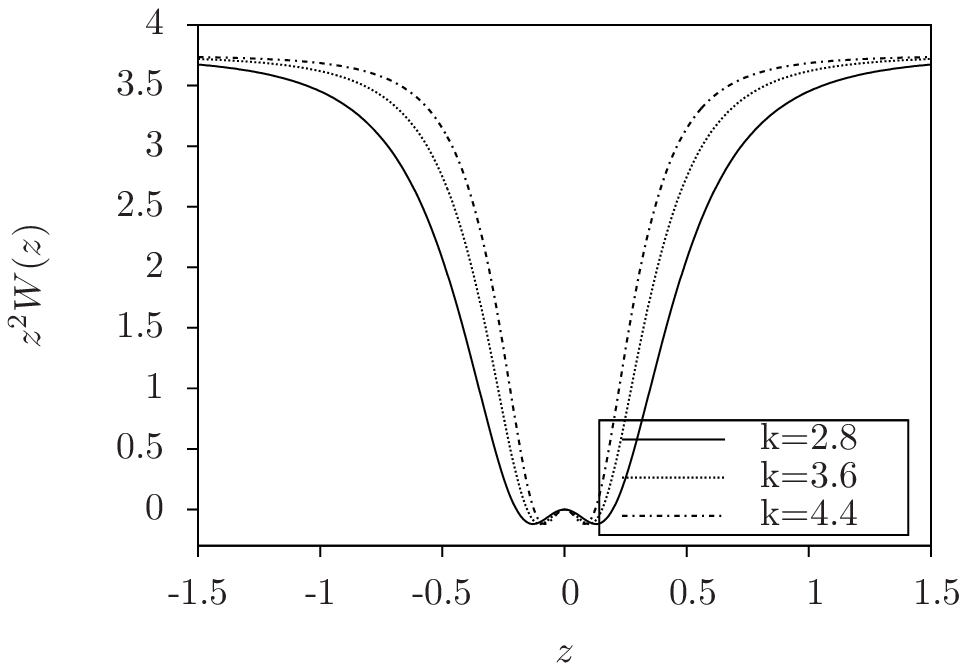}
%\end{center}
\caption{Plots of $W(z)$ and $z^2 W(z)$.}
\label{figureZzGravityPotential}
\end{figure}

Starting from $m^2>0$, the continuum of KK modes might lead to a
correction to the Newtonian potential on the brane. As depicted in
figure~\ref{figureZzGravityPotential}, $W(z)\sim\frac{15}{4z^2}$ as
$|z|\gg1$. It is well known (see, for
example~\cite{Csaki2000a,Bazeia2009}) that if the potential
$W(z)\sim\alpha(\alpha+1)/z^2$ as $|z|\gg1$, then the KK modes on
the brane take the form $\psi^{(m)}(0)\sim m^{\alpha-1}$, and for
two massive objects at a distance $r$ from each other, the
correction for the Newtonian potential is $\Delta
U\propto1/r^{2\alpha}$. For our model $\alpha=3/2$, thus
$|\psi^{(m)}(0)|^2\sim m$ for small masses. Thus the correction to
the Newtonian potential is $\Delta U\propto1/r^3$, which is same to the
one given by the RS model~\cite{Randall1999}.

%\begin{figure}
%\begin{center}
%\includegraphics[]{figureWarpFactor.eps}
%\end{center}
%\caption{Plots of the warp factor $e^{2A(y)}$. }
%\label{figureWarpFactor}
%\end{figure}
%
%
%\begin{figure}[h]
%\begin{center}
%\includegraphics[]{figurePhi.eps}
%\end{center}
%\caption{Plots of the scalar field $\phi(y)$ with $\kappa_5^2=1$. }
%\label{figurePhi}
%\end{figure}
%\begin{figure}[h]
%\begin{center}
%\includegraphics[]{figureScalarPotential.eps}
%\end{center}
%\caption{Plots of the scalar potential $V(\phi)$ for
%$\kappa_5^2=1$.} \label{FigureScalarPotential}
%\end{figure}
%
%\begin{figure}[h]
%\begin{center}
%\includegraphics[]{figureEnergyDensity.eps}
%\end{center}
%\caption{Plots of the energy density $\rho(y)$ by setting
%$\kappa_5^2=1$.} \label{figureEnergyDensity}
%\end{figure}

%\begin{figure}[h]
%\begin{center}
%\includegraphics[]{FigureGravityPotential.eps}
%\end{center}
%\caption{Plots of the Schrodinger potential $W(z)$.}
%\label{FigureGravityPotential}
%\end{figure}

Note that for KK modes with $0<m^2<W_{\rm{max}}$, there might exist
some resonant modes due to the quantum tunneling
effect~\cite{Gremm2000a,Csaki2000a}. Here $W_{\rm{max}}$ is the
maximum of the potential $W(z)$. To discuss this issue, let us
redefine the variables as
\begin{eqnarray}
\bar{z}=kz,\quad\bar{m}^2=m^2/k^2.
\end{eqnarray}
As a consequence, one can rewrite equation \eqref{Schrodinger} as
\begin{eqnarray}
  \left[\partial_{\bar{z}}^2
      -\frac{15 \left(-14+37 \bar{z}^2+28\bar{z}^4+4 \bar{z}^6\right)}{4 \left(5+7 \bar{z}^2+2 \bar{z}^4\right)^2}\right]
      \psi(\bar{z})
      =-\bar{m}^2\psi({\bar{z}}),
      \label{3.12}
\end{eqnarray} which is parameter independent. This equation is
hardly to be solved analytically, but we can apply some numerical
methods to analyze the resonant structure.

To start with, let us impose some initial conditions to simplify the
numerical
calculation~\cite{Liu2009a,Liu2009f,LiLiuZhaoGuo2011,Zhao2010c}:
\begin{eqnarray}
\psi_{\rm{even}}=c_1,\quad \psi'_{\rm{even}}=0;\quad
\psi_{\rm{odd}}=0,\quad \psi'_{\rm{odd}}=c_2.
\end{eqnarray}
Here we have denoted $\psi_{\rm{even}},~\psi_{\rm{odd}}$ as the even
and odd parity modes of $\psi(\bar{z})$, respectively. The prime
represents the derivative with respect to $\bar{z}$. In principle,
$c_1$ and $c_2$ are arbitrary constants, here we simply take $c_1=c_2=1$.
The massive modes $\psi({\bar{z}})$ here are analogous to the wave
function in the quantum mechanics. Therefore we can interpret
$|\psi({\bar{z}})|^2 d\bar{z}$ (after normalizing $\psi({\bar{z}})$), as the
probability for finding the massive KK modes within $(\bar{z},\bar{z}+d\bar{z})$. However,
the massive modes are non-normalizable because they oscillate
strongly when $\bar{z}\gg 1$. Following the method given
in~\cite{Liu2009a,Liu2009f,LiLiuZhaoGuo2011,Zhao2010c}, we define
the function
\begin{eqnarray}
P_{\rm{G}}(\bar{m}^2)=\frac{\int^{z_b}_{-z_b}|\psi({\bar{z}})|^2d\bar{z}}
{\int^{z_{\rm{max}}}_{-z_{\rm{max}}}|\psi({\bar{z}})|^2d\bar{z}}
\end{eqnarray}
as the relative probability for finding a massive KK mode in a
narrow range $-z_b < z<z_b$ (as compared to a wider interval
$-z_{\rm{max}} <z<z_{\rm{max}}$) with mass square $\bar{m}^2$. Here
$2z_b$ is about the width of the thick brane and
$z_{\rm{max}}=10z_b$. For an eigenvalue $\bar{m}^2\gg W_{\rm{max}}$,
the corresponding KK mode $\psi({\bar{z}})$ can be approximately
identified as a plane wave, for which the probability
$P_{\rm{G}}(\bar{m}^2)$ tends to $0.1~$.

\begin{figure}[h]
\begin{center}
\includegraphics[]{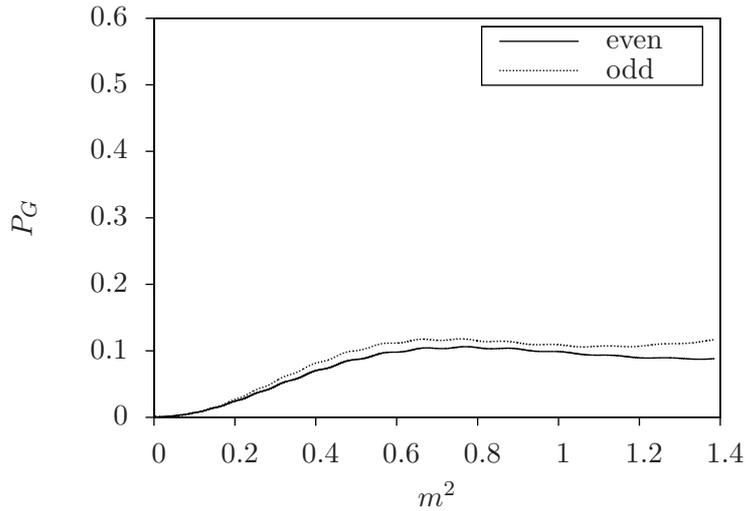}
\end{center}
\caption{The relative probability $P_{\rm{G}}$ for various values of
$\bar{m}^2$.} \label{figureProbabilityGravity}
\end{figure}

According to the numerical calculation of $P_{\rm{G}}(\bar{m}^2)$,
see figure~\ref{figureProbabilityGravity}, there exists no resonant
state for the Schr\"odinger equation \eqref{3.12}. Obviously, as the
eigenvalue $\bar{m}^2$ approaches to the maximum of the scaled
potential $\bar{W}_{\rm{max}}(\bar{z})=1.1086$, the relative
probability $P_{\rm{G}}$ tends to $0.1$, as expected.

 The analysis above indicates that the solution
(\ref{solutionpart1})-(\ref{solutionpart3}) is stable under tensor perturbations, and the
four-dimensional massless gravity can be localized on the brane.
Besides, there exists no resonant gravitational KK modes. Thus the
massive gravitons can not be quasi-localized on the brane. The
continuous KK spectrum contributes a correction to the brane
Newtonian potential, i.e., $\Delta U\propto r^{-3}$. Now let us move
on to the issue of trapping matter fields.

\section{Trapping of matter fields}
\label{section4}

An interesting but also vitally important question in brane world
scenarios is how to localize various bulk matter fields on the
brane by a natural mechanism. In many braneworld models the massless
scalar fields can be trapped on the
branes~\cite{Bajc2000,Liu2008c,Liu2010a}. Usually, Spin-1 Abelian
vector fields can not be localized on five-dimensional flat branes.
But, it can be localized on the RS brane in higher dimensional
case~\cite{Oda2000} or on the thick dS brane~\cite{Liu2009c} and
Weyl thick brane~\cite{Liu2008c,Liu2008}.

As far as our solution~\eqref{A_z} is concerned, the scalar fields
can be localized on the brane, while the vector fields can not. This
conclusion can be achieved by following the discussions given
in~\cite{Bajc2000,Liu2008}.

In addition to the scalar and vector fields, the localization of
Kalb-Ramond fields and gauge fields on single- or multi-branes
were also investigated~\cite{MukhopadhyayaSenSenGupta2009,TahimCruzAlmeida2009,ChristiansenCunhaTahim2010,LandimAlencarTahimGomesFilho2010,LiuFuGuoLi,Fu2011,DubovskyRubakov2001,Guerrero2010}. But we are not going to discuss these issues. In stead, we would
like to focus on the localization of the spin-1/2 fermions.

According to the works in the past, fermions can not be localized on
branes in five and six dimensions without introducing the
scalar-fermion
coupling~\cite{Ringeval2002,Melfo2006,Liu2007,Almeida2009,Liu2009,Liu2009a,Zhao2010b,Zhao2010c,ZhaoLiuWangLi2011,LiangDuan2009a}.
This is the reason why we insist on introducing a background scalar
field.

We proceed with the action
\begin{eqnarray}
 S_{\Psi} &=& \int d^4xdy \sqrt{-g} \bigg(\bar{\Psi} \Gamma^M (\partial_M + \omega_M) \Psi-\eta \bar{\Psi} F(\phi) \Psi\bigg),
 \label{DiracAction}
\end{eqnarray}
where $F(\phi)$ determines the coupling type and $\eta$ tells the
strength. As in refs.~\cite{Liu2008a,Liu2008,Liu2010}, we denote
$\Gamma^M=e^{-A}(\gamma^{\mu},~\gamma^5)$ as the five-dimensional
gamma matrices with $\gamma^{\mu}$ and $\gamma^5$ the usual
four-dimensional Dirac gamma matrices. $\omega_M$ are the spin
connections, of which the non-vanished components read $\omega_\mu
=\frac{1}{2}(\partial_{z}A) \gamma_\mu \gamma_5$. With the action in
\eqref{DiracAction}, we obtain the following Dirac
equation~\cite{Liu2008a,Liu2009,LiangDuan2009a,Zhao2010b}:
\begin{eqnarray}
 \left\{ \gamma^{\mu}\partial_{\mu}
         + \gamma^5 \left(\partial_z  +2 \partial_{z} A \right)
         -\eta¡¡e^A F(\phi)
 \right \} \Psi =0, \label{DiracEq1}
\end{eqnarray}
with $\gamma^{\mu}\partial_{\mu}$ the 4-dimensional Dirac operator.

It is convenient to decompose the full 5-dimensional spinor by
\begin{eqnarray}
\Psi(x,z) =\sum_n\psi_{\textrm{L}n}(x) f_{\textrm{L}n}(z)
 +\sum_n\psi_{\textrm{R}n}(x) f_{\textrm{R}n}(z).
\end{eqnarray}
Here $\psi_{\textrm{L}n}(x)$ and $\psi_{\textrm{R}n}(x)$ are the
left-hand and right-hand components of a 4-dimensional Dirac field,
which satisfy the four-dimensional Dirac equations
\begin{eqnarray}
\gamma^{\mu}\partial_{\mu}\psi_{\textrm{L}n}(x)=m_n\psi_{\textrm{R}n}(x),\quad
\gamma^{\mu}\partial_{\mu}\psi_{\textrm{R}n}(x)=m_n\psi_{\textrm{L}n}(x).
\end{eqnarray}
Now, we are left with the following coupled equations
\begin{subequations}\label{CoupleEq1}
\begin{eqnarray}
 \left[\partial_z + \eta\;e^A F(\phi) \right]f_{\textrm{L}n}(z)
  &=& m_n f_{\textrm{R}n}(z), \label{CoupleEq1a}  \\
 \left[\partial_z- \eta\;e^A F(\phi) \right]f_{\textrm{R}n}(z)
  &=&  -m_n f_{\textrm{L}n}(z). \label{CoupleEq1b}
\end{eqnarray}
\end{subequations}
Provided that eqs.~(\ref{CoupleEq1a})-(\ref{CoupleEq1b}), along with
the following orthonormal conditions
\begin{eqnarray}
\int_{-\infty}^{\infty}e^{4A} f_{\textrm{L}m}(z)
f_{\textrm{R}n}(z)dz =\delta_{\textrm{RL}}\delta_{mn},
\end{eqnarray} are satisfied,
we can recast action (\ref{DiracAction}) into the standard
four-dimensional actions for the massless and a series of massive chiral fermions. By
redefining $\tilde{f}_{\textrm{L}n}=e^{2A} f_{\textrm{L}n}$, we can
rewrite eqs.~\eqref{CoupleEq1} as two Schr\"{o}dinger equations
\begin{subequations}
\label{SchEqFermion}
\begin{eqnarray}
  \big(-\partial^2_z + V_{\rm{L}}(z) \big)\tilde{f}_{\textrm{L}n}
            &=&m_n^2 \tilde{f}_{\textrm{L}n},
   \label{SchEqLeftFermion}  \\
  \big(-\partial^2_z + V_{\textrm{R}}(z) \big)\tilde{f}_{\textrm{R}n}
            &=&m_n^2 \tilde{f}_{\textrm{R}n}.
   \label{SchEqRightFermion}
\end{eqnarray}
\end{subequations}
The effective potential for $\tilde{f}_{\textrm{L}n}$ has been
denoted as
\begin{eqnarray}\label{Vfermion}
  V_{\rm{L}}(y)&=&  e^{2A}\eta^2 F^2(\phi)
     -e^{A}\eta\partial_z F(\phi)
     -(\partial_z A)e^{A}\eta F(\phi)
\label{VL}.
\end{eqnarray}
For $\tilde{f}_{\textrm{R}n}$, the coresponding potential
$V_{\textrm{R}}(y)$ can be obtained from eq.~\eqref{VL} by simply
changing $\eta \to -\eta$. Note that in $z$-coordinate (defined by
eq.~\eqref{metricz})
\begin{eqnarray}
\phi(z)&=&\pm\frac{ \sqrt{2\lambda ^{(5)}}v^2 z }{\sqrt{3+2\lambda
^{(5)} v^2 z^2 }},\\
 A(z)&=&-\frac{1}{2} \ln \left(1+\frac{2}{3}\lambda ^{(5)} v^2 z^2
 \right).
\end{eqnarray}
Thus if we want to obtain $Z_2$ symmetric potentials $V_{\rm{L}}(y)$
and $V_{\rm{R}}(y)$, we have to constrain $F(\phi)$ to be an odd
function of $\phi$.

Now, our task is to solve the Schr\"odinger-like equation
\eqref{SchEqFermion} with different forms of $F(\phi)$ which
determines the shapes of the potentials $V_{\rm{L}}$ and
$V_{\textrm{R}}$. In braneworld models we mainly interested in two
types of effective potentials.

The first one is the so called volcano potential, which has been
discussed in section \ref{section3} (see eq.~\eqref{3.11}). For such
potential we usually get a unique bound state, i.e., the zero mode,
and a gapless continuum of KK modes. Sometimes, we might obtain
several quasi-localized massive states (the resonant states). In
section \ref{section3}, we got no sign which indicates the existence
of a resonant state. But as we will state below, such resonant
states might exists for particular $F(\phi)$.

The other one is the well known P\"{o}schl-Teller potential, which
approaches to a positive constant rather than zero as
$z\to\pm\infty$. This potential is very interesting, because in
addition to the zero mode, there might be some massive KK states.
Besides, the continuous KK modes are separated from the bound
states by a mass gap.

To proceed on, let us consider the following two cases, the first
one leads to a volcano potential, while the other one a
P\"{o}schl-Teller potential. Due to the fact that left- and
right-handed fermions usually share similar resonant and
massive spectrum~\cite{Zhao2010c}, we only discuss the left-handed
fermions.

\subsection{Massive fermions as resonant states}
By takeing $F(\phi)=\sqrt{\frac{3}{ 2\lambda ^{(5)}}}\phi$ we obtain
a Schr\"odinger equation with the following potential
\begin{eqnarray}
V_{\rm{L}}(z)&=&\frac{3 v^2 \eta  \left(v^2 z^2 \left(3 \eta +2
\lambda ^{(5)}\right)-3\right)}{\left(3+2 v^2 z^2 \lambda
^{(5)}\right)^2}\nonumber\\
&=&\frac{147 \eta  \left(147 \eta  z^2+29 \left(k^2 z^2-1\right)
\kappa _5^2\right)}{841 \left(1+k^2 z^2\right)^2 \kappa _5^4},
\label{volcanopot}
\end{eqnarray}
which satisfies $V_{\rm{L}}(0)=-v^2 \eta $ and
$V_{\rm{L}}(\infty)=0$. That means only the left-hand fermion zero
mode could be trapped on the brane for positive coupling constant $\eta$. The zero mode, according to the
discussions in~\cite{Liu2008a,LiangDuan2009a}, takes the form
\begin{eqnarray}
\tilde{\alpha}_{\textrm{L}0}(z)&\propto&\exp\left[-\eta\int^z_0 dz' e^{A(z')}\phi(z')\right]\nonumber\\
&=&\left(3+2\lambda ^{(5)}v^2z^2\right)^{-\frac{\beta}{2}},
\end{eqnarray}
where $\beta =\frac32\frac{\eta }{\lambda ^{(5)}}$ is a
dimensionless constant. %The integral
%\begin{eqnarray}
%&&\int_{-\infty}^{\infty}\left(3+2\lambda
%^{(5)}v^2z^2\right)^{-\beta}dz=3^{-\beta } \sqrt{\frac{3\pi }{2v^2
%\lambda^{(5)}}}\frac{ \Gamma(\beta -\frac{1}{2})}{ \Gamma(\beta)},
%\end{eqnarray} converges only when $\eta
%>\eta_0\equiv\lambda^{(5)}/3$.
%Here $\Gamma (\beta)=\int _0^{\infty }t^{\beta-1}e^{-t}dt$ is the
%Euler gamma function.
It is easy to proof that $\tilde{\alpha}_{\textrm{L}0}(z)$ is
normalizable if an only if $\eta
>\eta_0\equiv\lambda^{(5)}/3$. %That means if we regard $\lambda^{(5)}$ as a fundamental
%length scale, then the massless left chiral Fermion can be localized
%on the brane if and only if the coupling constant (as length) is far
%larger than $\lambda^{(5)}$.

\begin{figure}
%\begin{center}
\includegraphics[width=0.5\textwidth]{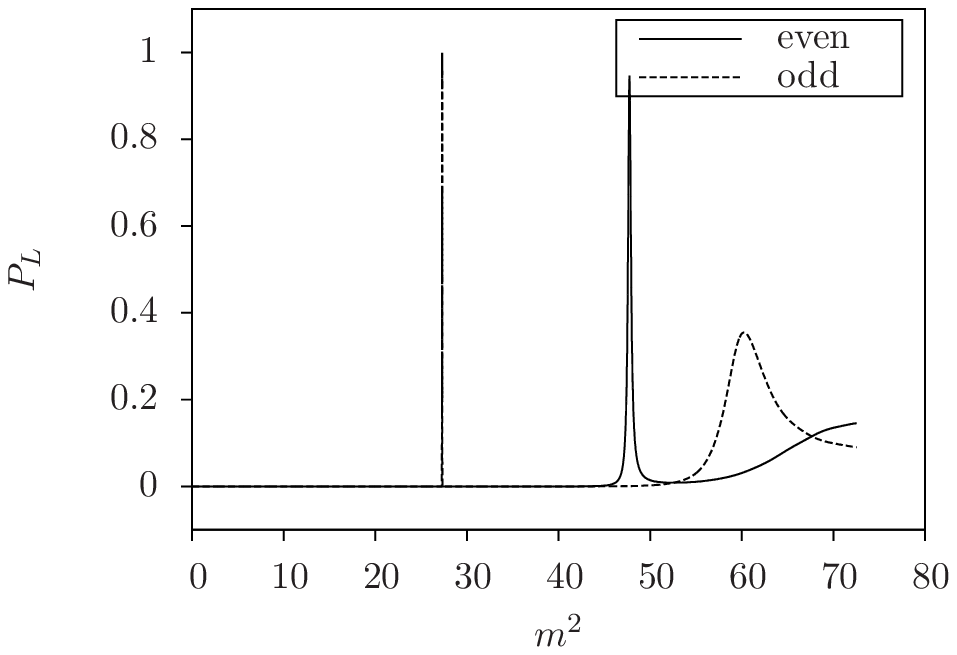}
\includegraphics[width=0.5\textwidth]{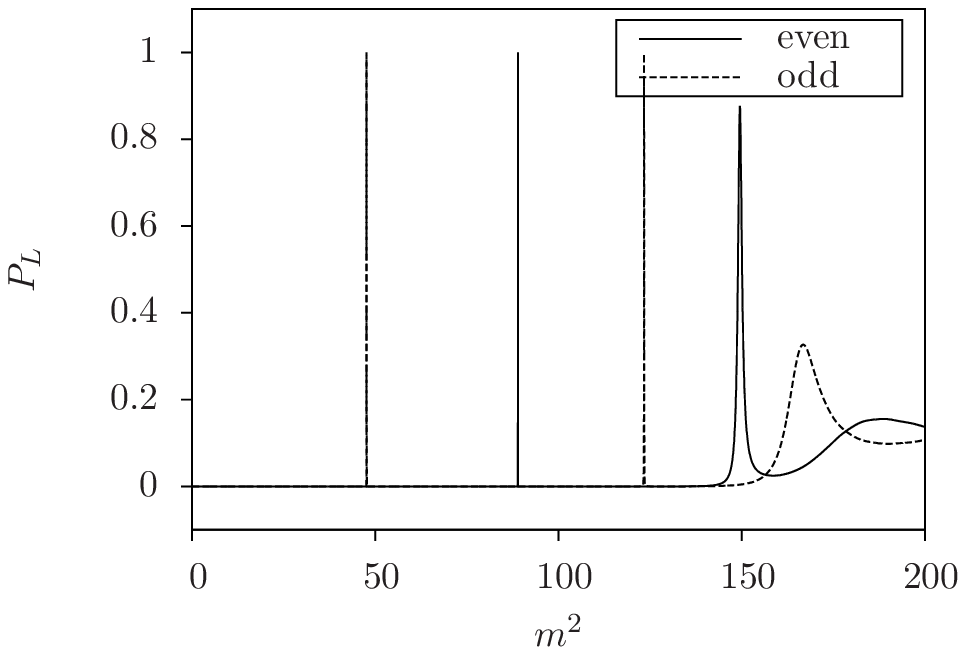}
%\end{center}
\caption{The figures of $P_{\rm{L}}$ with $k=1,~\kappa_5^2=1$,
$\eta=3$ (the left panel) and $\eta=5$ (the right panel).}
\label{figureRelaProbLeft}
\end{figure}

Since usually the fermions are massive, we hope to find a way to
trap the massive fermions. For volcano potential, the massive bound
state is absent. However, there might be some resonant states, which
have large relative probabilities near the location of the brane. The existence of
such resonant states offer us an effective way to trap massive
fermions. This matter trapping mechanism is refereed to as the
quasi-localization. Mimic to the case of gravity, we define
\begin{eqnarray}
P_{\rm{L},\rm{R}}(m^2)=\frac{\int^{z_b}_{-z_b}|\tilde{f}_{\rm{L},\rm{R}}({z})|^2dz}
{\int^{z_{\rm{max}}}_{-z_{\rm{max}}}|\tilde{f}_{\rm{L},\rm{R}}({z})|^2dz},
\end{eqnarray}
as the relative probability for finding a left- (or right-) hand
fermion resonant state $\tilde{f}_{\rm{L}}({z})$ (or
$\tilde{f}_{\rm{R}}({z})$) with mass $m$. Sharp peaks in figure of
$P_{\rm{L},\rm{R}}(m^2)$ can be identified as the resonant states.
The resonant states are unstable, their lifetime can be
estimated by $\tau\sim\Gamma^{-1}$, with $\Gamma=\delta m$ the width
of the half height of the resonant peak. As shown in
figure~\ref{figureRelaProbLeft}, for larger value of $\eta$, there
are more resonant states. The mass, width and
lifetime of the resonant states are listed in table~\ref{tab1}.

\begin{table}
\begin{center}
\begin{tabular}{|c|c|c|c|c|c|c|c|}
\hline
$\eta$      & $V_{\rm{L}}^{\rm{max}}$ & $n$  & $m^2$ & $m$ &$\Gamma$&$\tau$\\
\hline
 $3$         &  $58.033367$     &$1$& $27.28253$ & $5.22327$ & $0.000095$&10525.6787 \\
\cline{3-7}
  &        &$2$& $47.73727 $ & $6.90922$  & $0.0275038$&$36.3586$\\
\hline
       &   &$1$& $47.62055$ & $6.90076$ & $2.43773\times10^{-10}$&$4102176995.31678$\\
\cline{3-7}
$5$& $160.820587$ &$2$& $88.91114 $ & $9.42927 $  & $0.000005757$&$173695.37532$\\
\cline{3-7}
 &      &$3$& $123.32123$ & $11.10501$  & $0.0017165$&$582.58201$\\
\cline{3-7}
&   &$4$& $149.47803 $ & $12.22612$  & $0.0540928$&$18.48673$\\
\hline
\end{tabular}
\end{center}
\caption{The mass, width, and lifetime of resonances of the
left--handed fermions (shown in figure~\ref{figureRelaProbLeft}).
The parameters are $k=1$, $\kappa_5^2=1$. $V_{\rm{L}}^{\rm{max}}$ is
the maximum of the height of the potential $V_{\rm{L}}$, and $n$ is
the order of resonant states corresponding $m^2$ from small to
large.} \label{tab1}
\end{table}

\subsection{Massive fermions as bound states}
A coupling given by
\begin{eqnarray}
F(\phi)= k^2 \sqrt{\frac{v^2}{v^2-\phi ^2}}
\tanh\left[\frac{\sqrt{\frac{3}{2}} k \phi }{\sqrt{v^2 \lambda
^{(5)}\left(v^2-\phi ^2\right)}}\right]
\end{eqnarray}
would lead us to the P\"{o}schl-Teller potential
\begin{eqnarray}
  V_{\rm{L}}(z)= k^2\left[(k \eta)^2 -k \eta (1+k \eta ) \textrm{sech}^2(k z)\right],
  \label{fermionII}
\end{eqnarray}
for which $V_{\rm{L}}(0)=-k^3\eta,$ and $V_{\rm{L}}(\infty)=k^4\eta
^2$ is a positive constant. Again, the zero mode exists only when
$\eta>0$:
\begin{eqnarray}
\tilde{\alpha}_{\textrm{L}0}(z)=N_0\cosh^{-k\eta }(k z).
\end{eqnarray}
$\tilde{\alpha}_{\textrm{L}0}(z)$ is normalizable, provide that
$k\eta
>1/2$. For simplicity, let us take $k\eta=3/2$\footnote{In fact, when $k\eta=1,2,...$ is a positive integer, the quantum system possesses very peculiar property, namely, the system is reflectionless (and such property turns out to be intimately related to nonlinear integrable KdV system). Such peculiar property of Poschl-Teller system $V_L$ is reflected in the presence of a hidden nonlinear bosonized supersymmetry of order $2k\eta+1$ in it, see~\cite{CorreaPlyushchay2007,CorreaJakubskyPlyushchay2009}. We would like to thank Mikhail S. Plyushchay for reminding us of their interesting works.}, then the potential (\ref{fermionII}) reduces to
 \begin{eqnarray}
V_{\rm{L}}(z)=\frac{9}{4}k^2-\frac{15}{4} k^2\textrm{sech}(k z)^2.
\end{eqnarray}
As we have studied in~\cite{Liu2009,Liu2010b}, such potential
supports two bound states:
 \begin{eqnarray}
\tilde{\alpha}_{\textrm{L}0}(z)\propto\cosh^{3/2}(kz)\quad
&\textrm{with}& \quad m^2_0 =0,\\
\tilde{\alpha}_{\textrm{L}1}(z)\propto\sinh(kz)\cosh^{3/2}(kz)\quad
&\textrm{with}& \quad m^2_1 =2k^2.
\end{eqnarray}
%$\tilde{\alpha}_{\textrm{L}0}(z)\propto\cosh^{3/2}(kz)$ and
%$\tilde{\alpha}_{\textrm{L}1}(z)\propto\sinh(kz)\cosh^{3/2}(kz)$
%with the eigenvalues $m^2_0 =0$ and $m^2_1 =2k^2 <9k^2/4$, respectively.
We got only one massive bound state mode, i.e.,
$\tilde{\alpha}_{\textrm{L}1}(z)$. However, as the product $k\eta$
increases, more and more massive bound states appear. The number of
the bound states equals to the integer part of
$k\eta$~\cite{Liu2008c}. In addition to the bound states, there is
also a continuum of KK states which start from $m^2=k^4\eta ^2$ and
asymptote to plane waves as $z\gg1$.

\section{Conclusions and prospect}
In this paper, we study a thick domain wall solution in squared curvature gravity.
Different from the Gauss-Bonnet gravity, the dynamical equation in our model
contains higher order derivatives. As hoped, the solution itself is smooth and
regular, and is stable under the gravitational tensor perturbations. The
localization of gravity and spin-$1/2$ fermions is shown to be possible for our
solution. For gravity, only the zero mode can be localized. While for the
left-handed fermions, in addition to the zero mode, some massive KK modes can also
be localized as bound states (with infinite life), or be quasi-localized as resonant
states (with finite life). The number of massive KK modes depends on the thickness
of the brane $k\sim\gamma^{-1}$, and the strength of the scalar-fermion coupling,
i.e., $\eta$. The KK spectrum of the right-handed fermions is similar to the one of
the left-handed fermions, except the absent of the zero mode.

A new feature of our model is that the curvature square term might influence the
thick of the wall via the relation $k\sim\gamma^{-1}$. As the temperature effects is
considered, the coefficient $\gamma$ might be a function of temperature $T$, i.e.,
$\gamma=\gamma(T)$. Then, the thickness of the brane can be tuned by temperature.
Some massive KK states of the fermions might appear or disappear, for $k$ is another
factor which influences the number of the KK states of the fermions. Besides, the
Mexican hat potential \eqref{scalarPotential} indicates that there might be a
spontaneous symmetry breaking, which leads to phase transition and finally the
formation of our four-dimensional domain wall. The quantitative calculation is a
non-trivial task, we hope some related works can be reported in the near future.

 \acknowledgments
The authors would like to thank Chun-E Fu, Heng Guo and Ke Yang for
helpful discussions. This work was supported by the Program for New
Century Excellent Talents in University, the Huo Ying-Dong Education
Foundation of Chinese Ministry of Education (No. 121106), the
National Natural Science Foundation of China (No. 11075065), the
Doctoral Program Foundation of Institutions of Higher Education of
China (No. 20090211110028), and the Natural Science Foundation of
Gansu Province, China (No. 096RJZA055). Z.H. Zhao was supported by the Scholarship Award for Excellent Doctoral Student granted by Ministry of Education.

%\bibliographystyle{JHEP}
%\bibliography{D:/jabref/library/articles/articles}% Produces the bibliography via BibTeX.

\end{document}